%% file: main.tex
\documentclass[
reprint,
bibnotes,
amsmath,amssymb,
aps,
prb,
]{revtex4-1}

\usepackage{tabularx}
\usepackage{graphicx}% Include figure files
\usepackage{dcolumn}% Align table columns on decimal point
\usepackage{bm}% bold math
\usepackage{color}% color text
\usepackage{hyperref}% add hypertext capabilities
\usepackage{cleveref}% Automates figure and section references
\usepackage{caption}
\usepackage{float}
\begin{document}
\title{Nanowire melting modes during the Solid-Liquid Phase Transition: Theory and Molecular Dynamics Simulations}
%\thanks{A footnote to the article title}%

\author{Kannan M. Ridings}
\email{k.ridings@auckland.ac.nz}
\affiliation{MacDiarmid Institute for Advanced Materials and Nanotechnology,
Department of Physics, University of Auckland, Auckland 1142, New Zealand}

\author{Shaun C. Hendy}
\email{shaun.hendy@toha.nz}
% \affiliation{MacDiarmid Institute for Advanced Materials and Nanotechnology,
% Department of Physics, University of Auckland, Auckland 1142, New Zealand}
% \affiliation{Te P\={u}naha Matatini, Department of Physics, University of Auckland, Auckland 1142, New Zealand}
\affiliation{Toha Foundry, Auckland, New Zealand}

\date{\today}% It is always \today, today,
             %  but any date may be explicitly specified
\begin{abstract}
Molecular dynamics simulation have shown that after initial surface melting, nanowires can melt via two mechanisms: an interface front moves towards the wire centre; the growth of an instability at the interface can cause the solid to pinch-off and breakup. By perturbing a capillary fluctuation model describing the interface kinetics, we show when each mechanism is preferred and compare the results to molecular dynamics simulation. A Plateau-Rayleigh-type of instability is found, and suggests longer nanowires will melt via a instability mechanism, whereas in shorter nanowires the melting front will move closer to the centre before the solid pinch-off can initiate. Simulations support this theory; preferred modes that destabilise the interface are proportional to the wire length, with longer nanowires preferring to pinch-off and melt; shorter wires have a more stable interface close to their melting temperature, and prefer to melt via an interface front that moves towards the wire centre.
\end{abstract}

\maketitle

\maketitle
%\tableofcontents
\input{Stb_Intro.tex}
\input{Stb_Model2.tex}

\input{Stb_CompDetails.tex}
\input{Stb_Results.tex}
\input{Stb_Discussion.tex}
\input{Stb_Conclusion.tex}
\input{Acknowledgements.tex}
\bibliographystyle{apsrev4-1}
\bibliography{stability_new}
\end{document}

%% file: Stb_Intro.tex
\section{Introduction}
Nanostructured objects have lower stability with respect to their molten phase due to large surface area to volume ratios \cite{wronski1967size,coombes1972melting,di1995maximum}. In the case of nanowires their stability has been studied at elevated temperatures both experimentally  \cite{toimil2004fragmentation,shin2007size,xu2018situ} and theoretically \cite{dutta2014silico,ridings2019surface} indicating the presence of Plateau-Rayleigh (PR) type of instabilities can cause a nanowire to neck and breakup into a chain of nanospheres. In fact, PR like instabilities have been used as a means of self-assembly of chains of nanospheres for several different initial geometries ranging from rings \cite{nguyen2012competition}, wires \cite{fowlkes2012parallel}, and thin films \cite{roberts2013directed,hartnett2017exploiting}.
PR theory generally predict that the wavelength $\lambda$ of the perturbations which cause a liquid wire to become unstable are proportional to the initial wire circumference (\textit{i.e.} wires becomes unstable when $\lambda_{\textrm{c}} > 2 \pi R_0$). Moreover, linear stability analysis predicts a preferred wavelength that will drive a liquid wire to breakup.
Much work has been done in regards to nanocluster stability during solid-liquid coexistence \cite{schebarchov2006superheating}, and the stability of liquid nanojets \cite{moseler2000formation,eggers2002dynamics} and nanocylinders \cite{zhao2019revisiting}, relatively few studies address the stability of nanowires close to the melting point.
\\
It was found that for finite-sized cylinders during phase coexistence, differences in curvature and fluctuations would lead to the formation of random breaches at the material interface, causing the growth of instabilities which lead to the melting of the solid \cite{wu2015self}. For finite-sized boxes, different crystal geometries could be realised by overcoming nucleation barriers, where a crystal nucleus surrounded by its own fluid could change from a slab geometry, to a cylinder, and then to a solid droplet. It suggests that the solid prefers metastable forms as the box approaches the freezing (or melting) density \cite{statt2015finite}. Recent work has studied the breakage of gold nanofilaments connecting two nanoparticles where the filaments connecting the two nanoparticles would break apart by Joule heating  \cite{wu2022molecular}. Moreover, it was observed that the temperature at the breakage point had a strong dependence on the filament width, and had a dependence on the length in some, but not all cases \cite{wu2022molecular}.
\\
The thermally induced breaking of nanowires becomes important when considering the role they play in devices that utilise nanowire networks. Heat can be generated in nanowire networks via current passing through the network, and as such can influence the morphology and breakup of the nanowires making up the network  \cite{song2014nanoscale, volk2015thermal}. This could be a hindrance for device stability, where it is important to understand the limitations of interconnecting materials like nanowires. 
\\
In this paper, we investigate the stability of metal nanowires as they approach their melting temperature for copper nanowires of varying lengths and radii. To describe the nanowire stability, we perturb a capillary fluctuation model that describes the kinetics of the solid-liquid interface. The model is then tested against molecular dynamics (MD) simulations, where it is found that longer nanowires are more unstable with respect to the melt.

%% file: Stb_Model2.tex
\section{Capillary Fluctuation Model}
\begin{figure}[htp]
	\centering
{\includegraphics[width=0.4\textwidth]{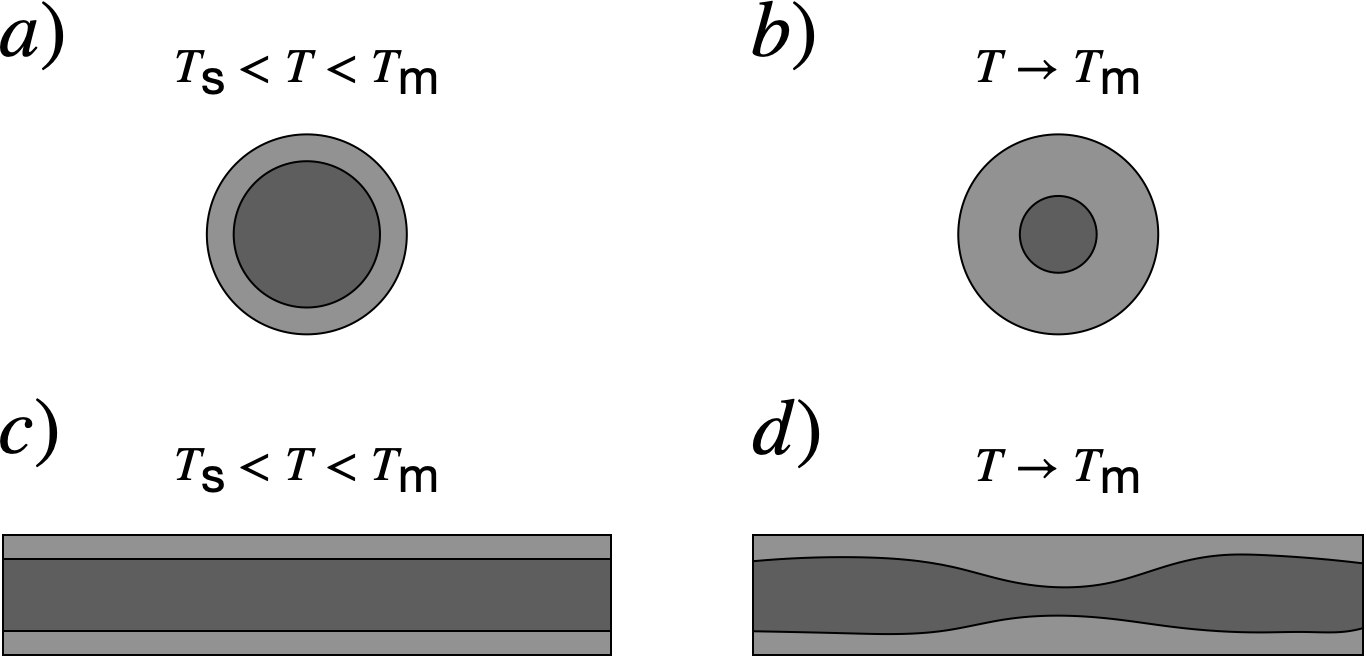}}
    \caption{Two different melting modes of nanowires where light grey represents the liquid, while the dark grey represents the solid. Panels a) to b) represent the moving interface front mechanism, whereas panels c) to d) show melting initiated via a instability at the solid-liquid interface.}
	\label{Fig:Melt_modes}   
\end{figure}

\begin{figure}[htp]
	\centering
{\includegraphics[width=0.5\textwidth]{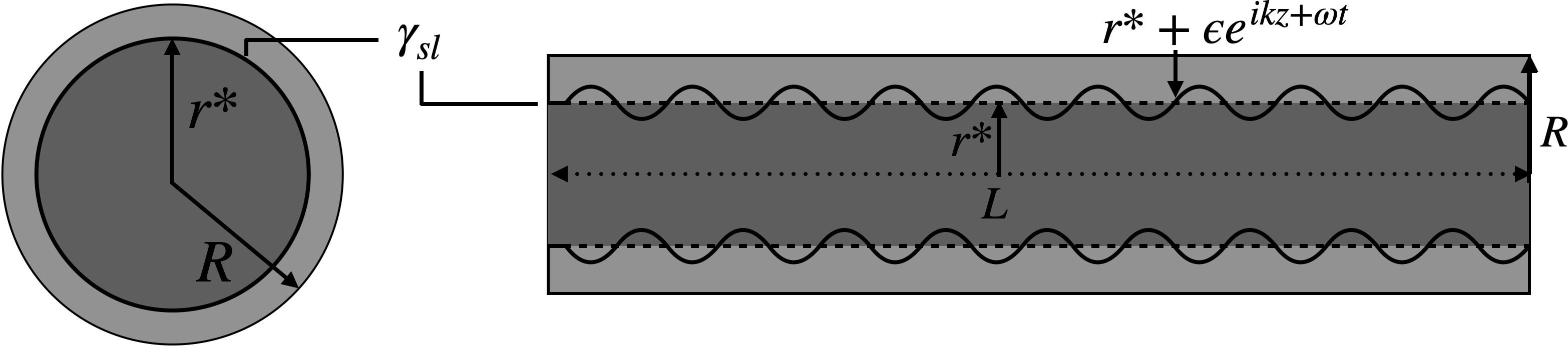}}
    \caption{A surface melted nanowire of radius $R_0$ and length $L$. The solid core radius during solid-liquid phase coexistence is $r^*$. This surface has undergone a small perturbation in the form of a plane wave $r^* + \epsilon e^{ikz + \omega t}$. $\gamma_{\textrm{sl}}$ represents the solid-liquid interfacial energy.}
	\label{Fig:Surf_Lin_melt}   
\end{figure}

Melting at the nanoscale is thought to initiate at the surface, and then move from the outside inwards, with the interface consuming the solid as it melts. However, it has been observed in nanowires that as $T\rightarrow T_{\textrm{m}}$ the solid will begin to neck and breakup \cite{ridings2019surface}. In Figure~\ref{Fig:Melt_modes} a) we see a top-down view a nanowire at a temperature $T$ that sits between its surface melting temperature $T_{\textrm{s}}$ and bulk melting temperature $T_{\textrm{m}}$. Figure~\ref{Fig:Melt_modes} b) shows as $T \rightarrow T_{\textrm{m}}$ the solid is consumed as the interface moves towards the wire centre. Figure~\ref{Fig:Melt_modes} c) shows a side-on view of a the same nanowire. However, as $T \rightarrow T_{\textrm{m}}$, rather than the interface moving towards the centre, a portion of the solid begins to thin out and neck, initiating the breakup of the solid as seen in Figure~\ref{Fig:Melt_modes} d).
\\
We construct our model by first considering the Gibbs free energy difference in an infinitely long cylinder surrounded by its own melt close to its melting temperature as dictated by classical nucleation theory \cite{wu2015self}
\begin{equation}
    \label{Eq:CNT_Cylinder}
\Delta G = - \pi r^2  L_{\textrm{v}}(\Delta T/T_c)+ 2 \pi r \gamma_{\textrm{sl}}.
\end{equation}
The value of $r$ that minimizes this equation represents the equilibrium solid radius for an infinitely long nanowire close to its melting point with solution 
\begin{center}
\begin{equation}
    \label{Eq:CNT_rstar}
r^* =  \frac{\gamma_{\textrm{sl}}T_c}{ L_{\textrm{v}} \Delta T}.
\end{equation}
\end{center}
Here, $\gamma_{\textrm{sl}}$ represents the solid-liquid interfacial energy, $L_{\textrm{v}}$ is the bulk latent heat of melting per unit volume, $T_c$ is the bulk melting temperature and $\Delta T$ is the undercooling. Now we wish to develop the idea of interface velocity for a cylindrical nucleus \cite{wu2015self}. For an infinite flat interface there is zero undercooling, and if $T<T_c$ then the solid-liquid interface will propagate towards the liquid phase with a velocity $V$ \cite{wilson1900xx, jackson1956kinetics,jackson1999computer,jackson2002interface}
\begin{equation}
    \label{eq:Jackson_Int_Vel}
    V = V_0\Big(1 - e^{-Q/k_b T}\Big),
\end{equation}
where $V_0$ represents a maximum velocity that depends on temperature, $Q$ is defined to be a thermodynamic driving force, and $k_b$ is Boltzmann's constant. This driving force is defined to be the difference between the solid and liquid phases per atom, so in a flat interface limit it can be approximated as $Q \simeq L_{\textrm{v}} \Delta T / N\, T_c$, where $N$ is the number density. Taking equation~\ref{eq:Jackson_Int_Vel}, substituting for $Q$ and taking $T = T_c - \Delta T$, then expanding in the small undercooling limit the interface velocity can be linearised as
\begin{equation}
    \label{eq:Int_Vel_Undercooling}
    V = \frac{V_0 L_{\textrm{v}}}{N k_b T_c^2}\Delta T = \zeta \Delta T,
\end{equation}
where $\zeta$ is a kinetic coefficient. This gives the planar interface velocity for the small undercooling limit. We now consider the kinetics of an interface by looking at the dynamic behaviour of an interface with a profile $r(z,t)$ \cite{hoyt2010fluctuations,wu2015self,wu2021crystal}
\begin{equation}
    \label{eq:Interface_Kinetics_r(z,t)}
    \frac{d\Tilde{r}}{dt} = \zeta \Gamma \frac{d^2\Tilde{r}}{dz^2} + \zeta \Tilde{\eta} + \zeta \Delta T\Big(1 - \frac{r^*}{\Tilde{r}}\Big),
\end{equation}
% This argument on interface velocities can be further extended by considering the kinetics of an interface by looking at the dynamic behaviour of an interface with a profile $h(x,t)$ \cite{hoyt2010fluctuations}
%     \begin{equation}
%     \label{eq:Interface_Kinetics}
%     \frac{dh}{dt} = \zeta \Gamma \frac{d^2h}{dx^2} + \zeta \Tilde{\eta} + \zeta \Delta T,
% \end{equation}
where $\Gamma = (\gamma_{\textrm{sl}} + \gamma_{\textrm{sl}}'')T_c/L_{\textrm{v}}$, with $\gamma_{\textrm{sl}} + \gamma_{\textrm{sl}}''$ the interfacial stiffness of the solid-liquid interface \cite{morris2002complete}. $\Tilde{\eta}$ is a thermal noise term similar to a Langevin-type description. This takes into account fluctuations about equilibrium and satisfies $\langle \Tilde{\eta}(x,t)\Tilde{\eta}^*(x',t')\rangle = \mathcal{C}\delta(x - x')\delta(t-t')$, where $\mathcal{C}$ is a constant and the delta functions suggest the noise is uncorrelated in space and time. 
% The solid-liquid interface has been observed as atomically rough in Lennard-Jones systems \cite{morris2003anisotropic}, hard-sphere \cite{mu2005anisotropic, davidchack2006anisotropic}, and metals and alloy systems \cite{asta2002calculation, potter2011molecular}. This is thought of to occur due to the presence of thermal fluctuations, which have are typically pronounced in solid-liquid interfaces due to their low surface energy. 
% Taking this interface profile $h(x,t)$ and expressing it in terms of a radial coordinate $\Tilde{r}(z,t)$ the following interface profile can be recovered
% \begin{equation}
%     \label{eq:Interface_Kinetics_r(z,t)}
%     \frac{d\Tilde{r}}{dt} = \zeta \Gamma \frac{d^2\Tilde{r}}{dz^2} + \zeta \Tilde{\eta} + \zeta \Delta T\Big(1 - \frac{r^*}{\Tilde{r}}\Big),
% \end{equation}
% where the undercooling term has been modified to account for the finite sized effects in the growth of the nucleus from $\Tilde{r}$ to $\Tilde{r} + d\Tilde{r}$, taking into account the addition of curvature into the system. 
We perturb the solid-liquid interface by a small parameter $\epsilon$, so we can express it as the surface $\Tilde{r}(z,t) = r^* + \epsilon e^{ikz + \omega t}$, as seen in Figure~\ref{Fig:Surf_Lin_melt}.
A stability analysis can be carried out on a system with interface kinetics governed by such an equation as~\ref{eq:Interface_Kinetics_r(z,t)}. If one looks at the surface in equilibrium $r^*$ and then perturbs that with a wave, that disturbance can be approximated as $\Tilde{r} = r^* + \epsilon e^{i k z + \omega t}$, with $\epsilon$ being a small perturbation, with $k$ and $\omega$ being the wavenumber and instability growth rate respectively. The term $(r^* + \epsilon e^{i k z + \omega t})^{-1}$ can be approximated as
\begin{equation}
    \label{eq:surface_expansion}
    \frac{1}{r^* + \epsilon e^{i k z + \omega t}} \simeq \frac{1}{r^*} - \frac{\epsilon}{r^{*2}}e^{i k z + \omega t}.
\end{equation}
By substituting $\Tilde{r}$ into equation~\ref{eq:Interface_Kinetics_r(z,t)}, using the expression in equation~\ref{eq:surface_expansion},  solving to $\mathcal{O}(\epsilon)$, assuming an isotropic solid-liquid interface (i.e. we can take $\Gamma \approx \frac{\gamma_{\textrm{sl}}T_c}{L_{\textrm{v}}}$ \cite{wu2021crystal}), and using the definition of $r^*$ in equation~\ref{Eq:CNT_rstar} we recover
\begin{equation}
    \label{Eq:omega}
    \omega = \frac{\zeta \Delta T}{r^*}\Big(1 - k^2 r^{*2} \Big).
\end{equation}
A PR type instability can be found by observing that $\omega > 0$ when $kr^* < 1$, bringing us to the familiar solution
\begin{equation}
    \label{Eq:PR_Stability}
\lambda^* >   2 \pi r^*.
\end{equation}
We combine equations~\ref{Eq:PR_Stability} and \ref{Eq:CNT_rstar} and take $\Delta T = T_c - T_{\textrm{m}}$, where $T_\textrm{m}$ is the bulk melting temperature of a nanowire of radius $R_0$. The interface will remain stable when $\omega < 0$ and leads  to the moving interface front seen in Figure~\ref{Fig:Melt_modes} a) to b) giving the condition
\begin{center}
\begin{equation}
    \label{Eq:T_omega}
T_{\omega} < T_{\textrm{m}},
\end{equation}
\end{center}
where we define $T_{\omega}=T_c\bigg(1 - \frac{2 \pi \gamma_{\textrm{sl}}}{ L_{\textrm{v}} \lambda^*}\bigg)$. For $\omega > 0$ we have $ T_{\omega} > T_{\textrm{m}}$, and perturbations will grow to destabilise the solid-liquid interface, giving the scenario in Figure~\ref{Fig:Melt_modes} c) to d). If $\lambda^* \propto L$ then $T_{\omega}$ becomes larger than $T_{\textrm{m}}$ quickly, giving a clear criteria for describing when each melting mode is preferred.
\\
Finally, by combining equations~\ref{Eq:CNT_rstar} and $T_{\textrm{m}}$ from previous work \cite{ridings2019surface}, a equation for the equilibrium solid radius $r^*$ in terms of initial wire radius $R_0$ and interfacial energies $\kappa$ (where $\kappa = \frac{1-\Delta \gamma /\gamma_{\textrm{sl}}}{1+\Delta \gamma /\gamma_{\textrm{sl}}}$ and $\Delta \gamma$ is the spreading parameter that determines the wettability of a material \cite{ridings2019surface})
\begin{center}
    \begin{equation}
    \label{Eq:rstar_R0}
    r^* = \frac{R_0}{4}\bigg( \kappa +  \frac{I_0(R_0/\xi)}{I_1(R_0/\xi)} \bigg),
\end{equation}
\end{center}
which shows how the equilibrium solid radius scales with size. 
% This result states there is a criteria to be met in order for unstable modes in a cylindrical solid-liquid interface to grow. The implication is once $\lambda > \lambda_{stab}$ then the unstable modes which will eventually initiate the pinch-off of the solid are enabled to grow in time. In this paper we will see that this is a crucial detail as to how solid-liquid interfaces behave in nanowires. 

%% file: Stb_CompDetails.tex
\section{Computational Details}
\label{S:Comp_Details}
In this section we outline the approach taken for our molecular dynamics simulations. Periodic boundary conditions along the wire axis were used to simulate an infinitely long wire, and additionally suppress long-wavelength instabilities that may otherwise cause the wire to break apart prior to completely melting. The FCC nanowires are bounded by \{100\} and \{110\} surfaces, which were made via a Wulff-type construction (see \cite{ridings2019surface} Figure 2).
\\
Interactions between atoms were modelled using an embedded-atom-model (EAM) potential for copper, developed by Sheng \textit{et al} \cite{sheng2011highly}. The bulk melting temperature for this copper potential is 1320 K. This value is slightly below the experimental bulk melting temperature which ordinarily occurs for EAM potentials. This potential was chosen because it makes good predictions of bulk properties and yields fairly good melting temperatures. A previous study with nickel yielded good results \cite{ridings2019surface} for melting temperatures and dynamics. 
\\
To account for the expansion of the lattice, an atomic volume of approximately $V_{\textrm{atom}}=13.2$\AA\ was used. This assumed the density of copper at the melting point was $\rho_{_{Cu}}=8020$ kg/m$^3$. 
\\
The equations of motion were integrated using a Verlet method using a timestep of 2.5 fs. To control the temperature a Langevin thermostat with a damping parameter of $1.0$ ps$^{-1}$. This was to ensure a quick equilibration at each timestep of the simulation. The simulations were initialized at an initial temperature $T_i$ for 1.0 ns. Afterwards, a production phase for each wire was run from a temperature $T_i$ to a temperature $T_i + 1$. Then an equilibration phase around the temperature $T = T_i + 1$ was run. Each production phase was 0.40 ns, with an equilibration phase of 0.60 ns, creating an effective heating rate of around $1$K/ns. This ensured us that at each temperature the wires were sufficiently close to equilibrium. 
\\
Molecular dynamics simulations were employed to test the theory developed in the previous section. We chose three nanowires with initial radii of 22.0 \AA, 30.0 \AA, and 38.0 \AA. For each nanowire, lengths were chosen so that the aspect ratios $L/R_0$ were all 5.0, 7.5, 15.0, and 25.0 were satisfied. These four wire lengths were chosen to ensure that $L_1 < \lambda_{\textrm{c}}$, $L_2 \simeq \lambda_{\textrm{c}}$, $L_3 > \lambda_{\textrm{c}}$ and $L_4 \gg \lambda_{\textrm{c}}$, where $\lambda_{\textrm{c}}$ represents the classical Plateau-Rayleigh critical wavelength (recall that perturbations grow when $\lambda_{\textrm{c}} > 2\pi R_0$). Each individual wire had 10 - 20 individual runs to gather statistics on the quantities of interest.
\\
To obtain values of $r^*$ from simulation, coordinates of the solid atoms at the solid-liquid interface were extracted using a nearest-neighbour type of algorithm. We averaged values of $r^*(z)$ just prior to the solid pinch-off along the length of each wire to and averaged it to estimate $\bar{r^*}$. To extract the modes which destabilised the solid-liquid interface, the Fourier transform of $r^*(z)$ was taken for each run on each individual wire. A sampling frequency $F_s = 4$ was taken, since this is the closest whole number to the lattice spacing of copper. The radius for the solid were normalized such that the mean would be zero. To allow a more reliable extraction of $k_{sol}$ ($k_{max}$ for the solid-liquid interface), peaks were resolved with zero padding to give a longer signal. Destabilising modes of $k_{sol}$ were then averaged over each run. This ensures a more reliable estimate of the Fourier profile for each individual wire.
\\
Finally, to estimate values of the melting temperature, we calculated heat capacity $C_{\textrm{v}} = \frac{dE}{dT}$ and used the peak to identify the bulk melting temperature of each wire. In each case, the heat capacity diverges when the wire begins to melt.

%% file: Stb_Results.tex
\section{Simulated Results}
\begin{figure}[htp]
	\centering
{\includegraphics[width=0.40\textwidth]{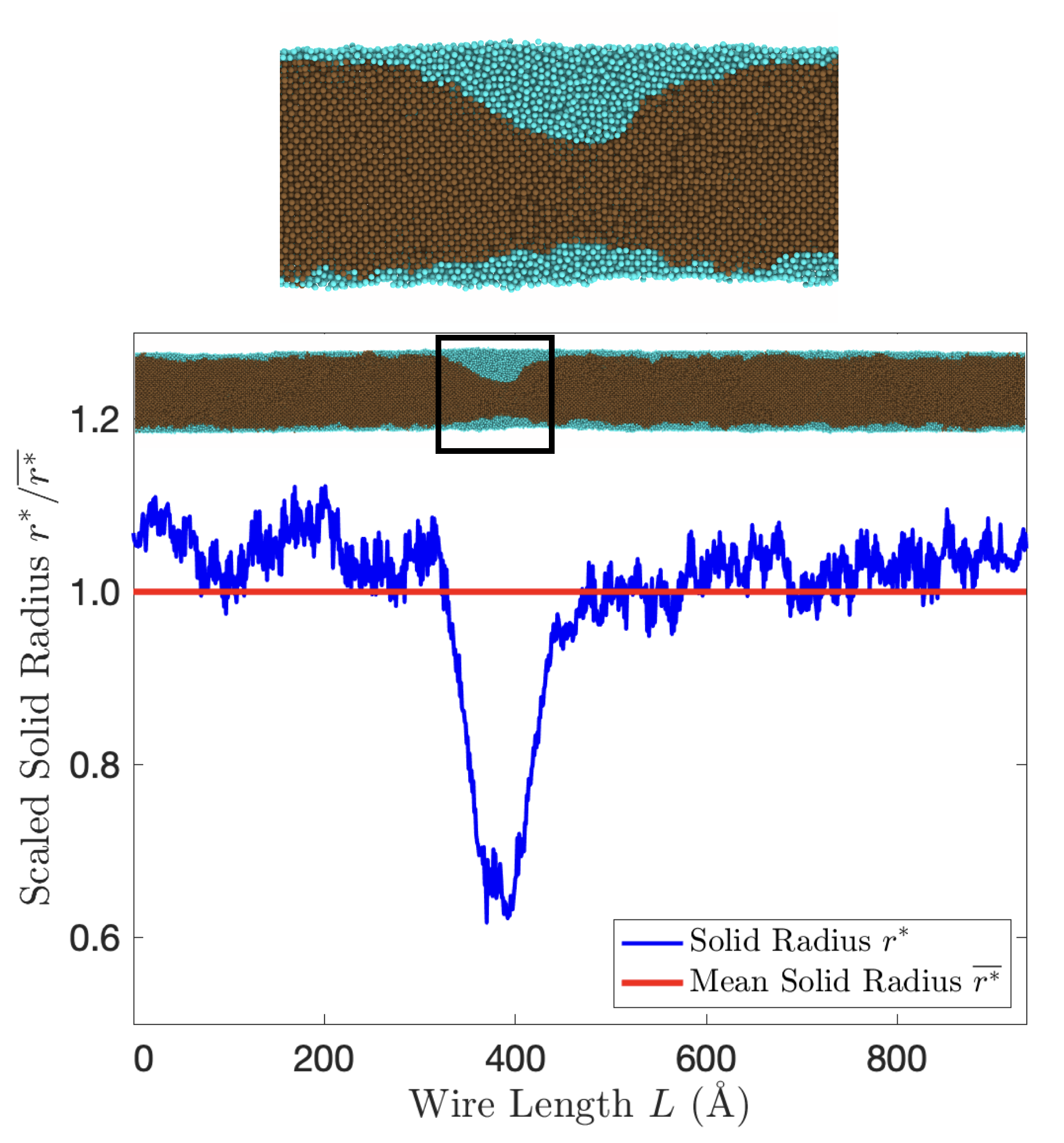}}
    \caption{A plot showing the solid radius $r^*(z)$ of a surface melted nanowire scaled by the average solid radius $\overline{r^*}$ just prior to the wire fully pinching off. The solid atoms are coloured brown (darker) and the liquid atoms are coloured blue (lighter). Accompanying the plot is a snapshot of this nanowire, and a zoomed in version of the solid pinch-off.}
	\label{Fig:rad_vs_z}   
\end{figure}

\begin{figure}[htp]
	\centering
{\includegraphics[width=0.45\textwidth]{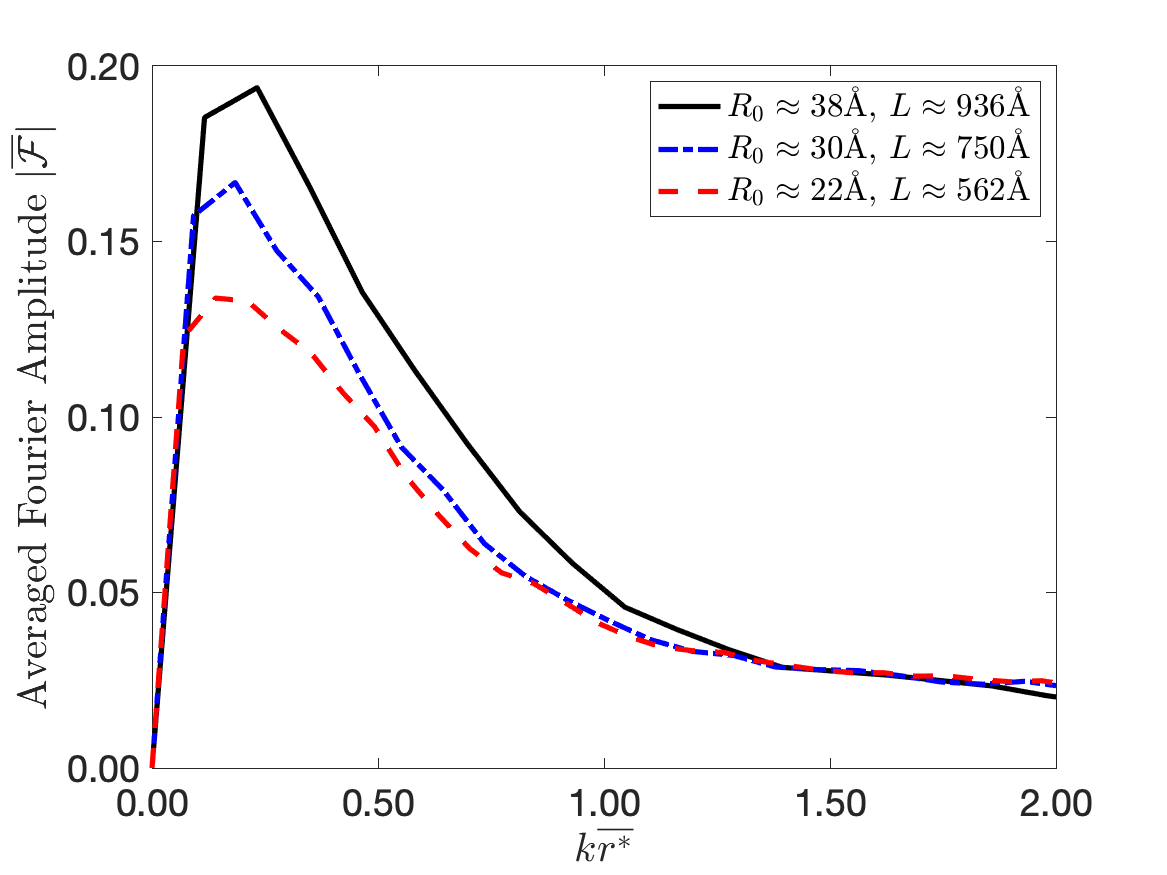}}
    \caption{The averaged Fourier transform of wires with $L/R_0 = 25$. Values of $k_{sol}$ for each wire are all low, indicating modes that destabilise the solid-liquid interface are proportional to the wire length.}
	\label{Fig:FT_L4}   
\end{figure}

\begin{figure}[htp]
	\centering
{\includegraphics[width=0.45\textwidth]{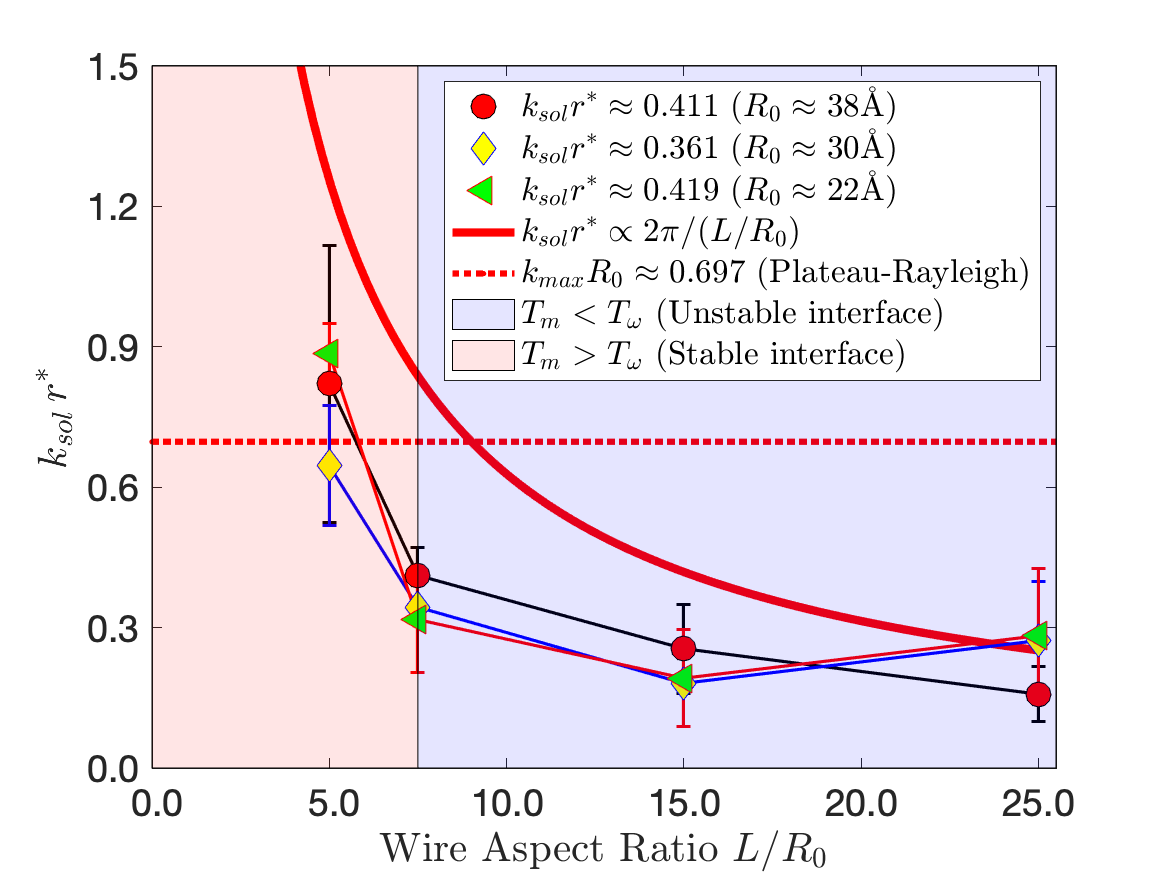}}
    \caption{This figure shows how the wire aspect ratio effects the fastest growing modes. A thick red line plots $2 \pi /(L/R_0)$, which we assume is proportional to $k_{sol}r^*$.}
	\label{Fig:kstar_solid}   
\end{figure}

We begin by first looking at the stability of the solid-liquid interface by examining the solid-core prior to the point where the solid pinches off, as shown in Figure~\ref{Fig:rad_vs_z}. The solid begins to neck at some point as the wire is heated close to its bulk melting temperature $T_{\textrm{m}}$. As we will see in this section, wires with lengths that satisfy $L > 2\pi R_0$ will pinch-off and melt at a temperature consistently lower than wires with lengths $L \leq 2\pi R_0$. 
\\
Figure~\ref{Fig:FT_L4} shows the Fourier transform for wires of initial radii $R_0 = 38,\,30,\,22\,$\AA, with each satisfying $L/R_0 = 25$. The values for $k_{sol} r^*$ are similar, while the values for $\overline{r^*}$ differ. 
\\
Figure~\ref{Fig:kstar_solid} represents a stability diagram in terms of the fastest growing modes $k_{sol}r^*$ against wire aspect ratio $L/R_0$. The modes $k_{sol}r^*$ for each wire aspect ratio are similar, which indicates destabilising modes depend strongly on the wire aspect ratio. As nanowires get shorter, modes that destabilise the interface approach unity, in violation of classical PR theory (see red dashed line). Also seen are two regimes which identify the preferred melting mode, as seen in Figure~\ref{Fig:Melt_modes}. To the left (red region) we see the regime where $T_{\omega} < T_{\textrm{m}}$ which indicates the solid-liquid interface must move closer to the centre before the pinch-off can initiate. On the right (blue region) we see the regime where $T_{\omega} > T_{\textrm{m}}$ and identifies when the pinch-off melting mechanism is favoured. Observations from MD simulation agree with the theory developed, represented by equations~\ref{Eq:omega}, ~\ref{Eq:PR_Stability} and ~\ref{Eq:T_omega}. Longer wires will be more thermodynamically unstable since wire lengths will generally be greater than the circumference of the coexisting solid, as indicated by equation~\ref{Eq:PR_Stability}. Included in Figure~\ref{Fig:kstar_solid} is the curve $k_{sol}r^* \propto \frac{2 \pi R_0}{L}$, which follows the trend observed in MD simulations, as well as predictions by classical PR theory which states $k_{max}R_0 \simeq 0.697$. For wires with $L<2\pi R_0$, $k_{sol}r^*$ approaches unity, violating classical PR theory, and indicating regions of interface stability.
\\
We now examine how the wire length influences the stability of the solid-liquid interface by looking at how $r^*$ obtained from simulation behaves close the the melting point for wires of different radii and lengths. From equation~\ref{Eq:CNT_rstar} we can see that $r^* \propto \Delta T^{-1}$.
\begin{figure}[htp]
	\centering
{\includegraphics[width=0.45\textwidth]{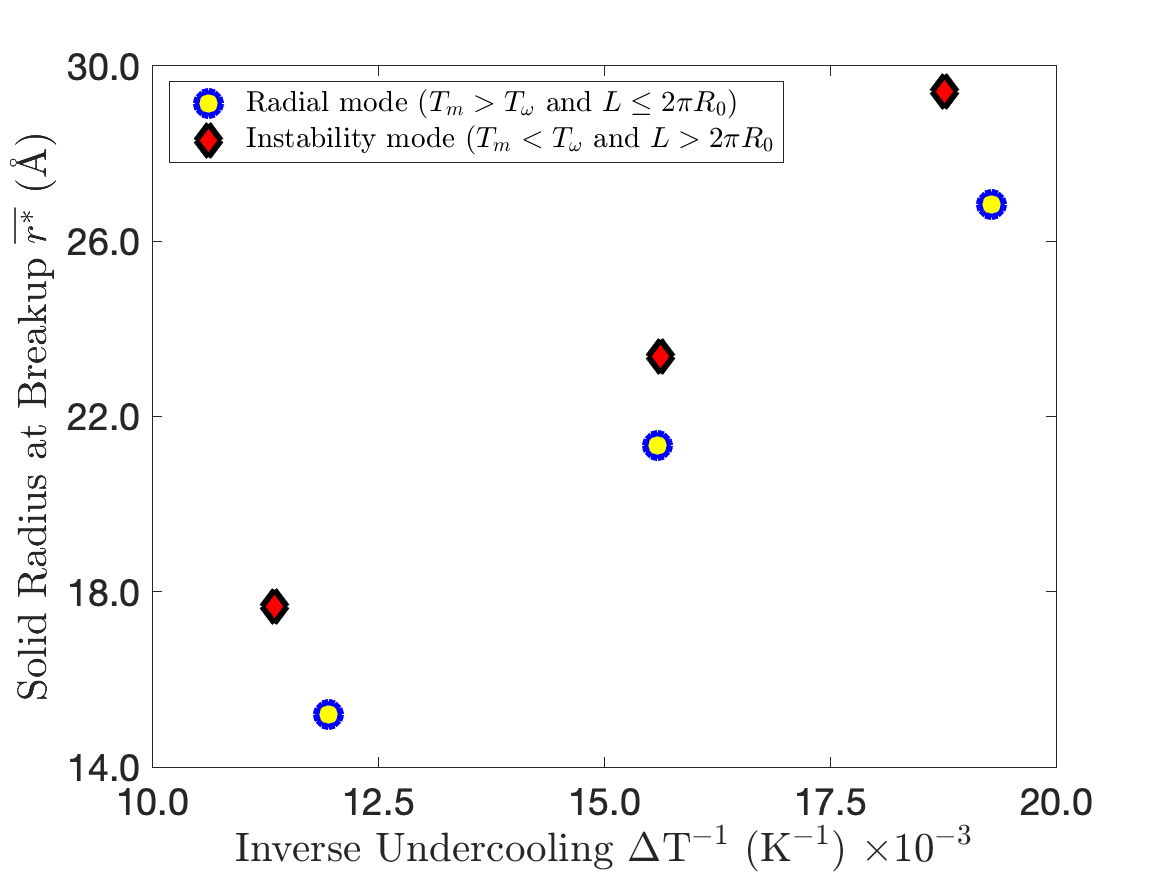}}
    \caption{Values of $r^*$ obtained from simulation against $\Delta T^{-1}$. Yellow circles are for $r^*$ when $L \leq 2\pi R_0$, and red diamond points represent $r^*$ and $\Delta T$ for when $L > 2\pi R_0$. Yellow circles indicate when the radial mode is the preferred melting mechanism, and red diamonds indicate where the instability mode is preferred. The first two points (bottom left) are for wires where $R_0 \approx 22$\AA\, the middle two are for $R_0 \approx 30$\AA, and the last two are for $R_0 \approx 38$\AA. We assume yellow and red points represent the bulk and instability melting temperatures, $T_{\textrm{m}}$ and $T_{\omega}$, respectively.}
	\label{Fig:rstar_DT}   
\end{figure}
As observed in Figure~\ref{Fig:rstar_DT}, MD results consistently show that longer wires melt at a lower temperature since they are prone to the growth of instabilities that initiate the pinch-off. Shorter wires not only melt at a slightly higher temperature, but the value of $r^*$ at the point when the pinch-off initiates is consistently smaller, in agreement with the theory developed. It supports the idea that there are two melting mechanisms that depend on the wire aspect ratio, as evident in Figure~\ref{Fig:kstar_solid}. The simulated results point to the bulk melting temperature of the potential used being about 1230 K, rather than the 1320 K stated. 
\begin{figure}[htp]
	\centering
{\includegraphics[width=0.45\textwidth]{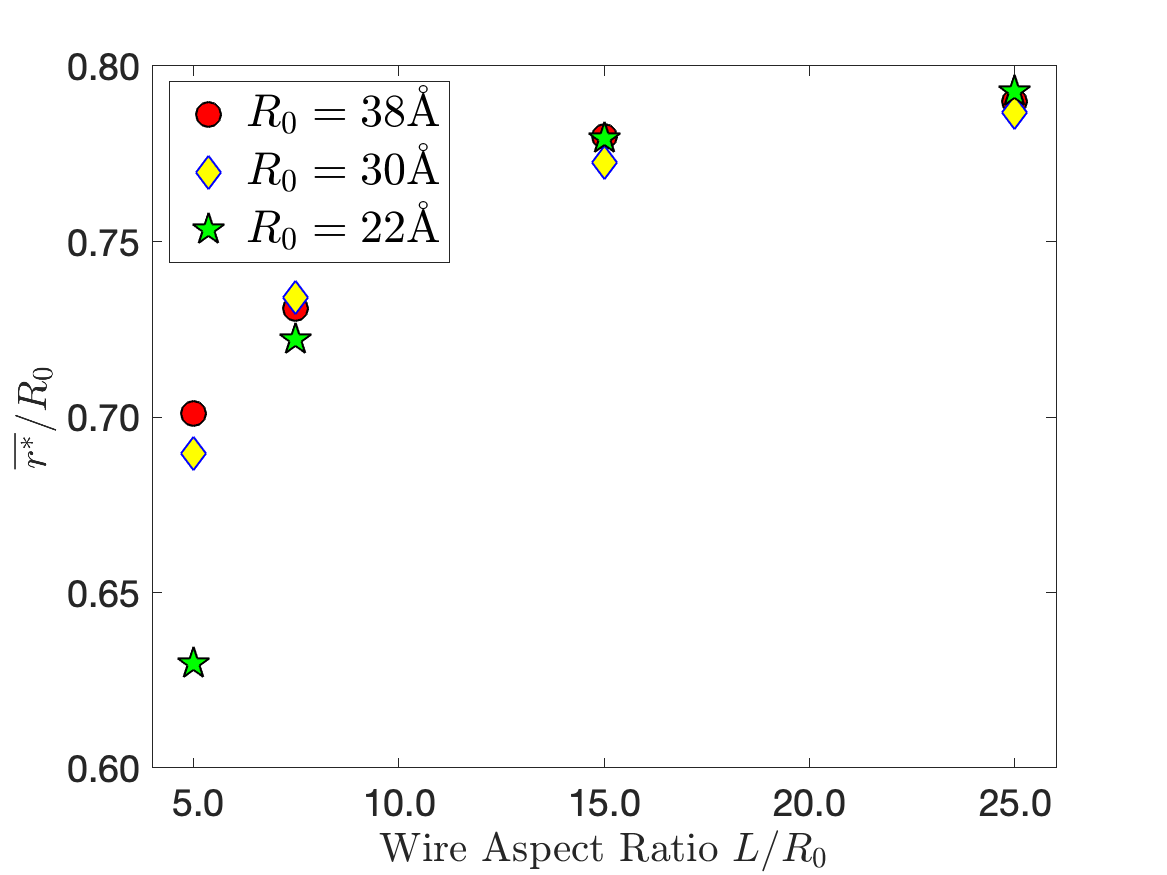}}
    \caption{The ratio of equilibrium solid radius to initial wire radius against the wire aspect ratio. Values of $r^*/R_0$ for each aspect ratio are clustered closely together, where they begin to deviate markedly when the initial wire length $L < 2 \pi R_0$.}
	\label{Fig:rstar_R0}   
\end{figure}
\\
In Figure~\ref{Fig:rstar_R0} we see that $r^*$ depends not only on the initial wire radii $R_0$, but also the wire aspect ratio as well. This evidence supports the theory and previous claims that as the wire aspect ratio gets smaller, the solid core radius prior to the pinch-off decreases. The ratio of $r^*/R_0$ appears to converge in the limit of large $L/R_0$. Each $r^*/R_0$ for the aspect ratios studied are similar when $L> 2\pi R_0$. Once $L \leq 2\pi R_0$ the difference in $r^*/R_0$ becomes appreciable. This could be indicative that quantities like the interfacial energies and correlation length ($\gamma_{\textrm{sl}}$ and $\xi$ respectively) become important quantities for small wires, implying size and curvature play key roles in observations for small aspect ratios.

%% file: Stb_Discussion.tex
\section{Discussion}
By perturbing the interface of the two-parabola model used in our previous work \cite{ridings2019surface} we can describe when melting will take place via a instability or radially. The model showed the fastest growing modes that destabilise the interface are inversely proportional to the wire length, where a PR type instability for the solid in a surface melted nanowire is recovered. By using classical nucleation theory and exploring the nanowire stability in the vicinity of $T_{\textrm{m}}$, we were able to define the condition that determines the preferred melting mechanism. Moreover, we recovered an expression for the equilibrium solid radius in terms of the initial wire radius and the interplay of the interfacial energies of the nanowire.
\\
Simulations show the fastest growing modes are inversely proportional to the wire length, and in fact that $k_{sol}r^* \propto \frac{2\pi R_0}{L}$. Additionally, we observe longer nanowires consistently melt at a lower temperature than shorter wires, in agreement with our developed theory and other recent observations \cite{wu2022molecular}. The implication is that shorter nanowires have a more stable interface when close to their melting temperature. In some cases it was observed that for the longest, thinnest nanowires, the liquid-vapour interface would begin to neck, being driven by surface diffusion, which in turn influenced the breakup of the solid. For longer heating rates or overdamped Brownian dynamics this feature would become more pronounced, but due to the quick equilibration at each timestep this was not an issue.
\\
Clear evidence can also be seen that the equilibrium solid radius depends on the wire aspect ratio, and not just the radius. It shows interface energies dependence on the nanowire surface area, rather than just their radius. Studies have explored the size-dependence of interfacial energies \cite{jiang2008size}. Curvature too, plays a role in the interfacial energy, where for spherical clusters the solid-liquid interface energy is linear with inverse radius \cite{montero2019interfacial, montero2020interfacial}. This size and curvature dependence explains why values of $r^*/R_0$ begin to deviate away small low aspect ratio. The ratio of atoms at the surface compared to the bulk becomes far more appreciable for the smallest wires, giving the curvature a greater role in the solid-liquid interface dynamics. Given the fact that the fastest growing modes are inversely proportional to the nanowire length, it would be of no surprise that interfacial energies will have dependence on this too, since their surface area will scale with radius and length. The theory and simulations show that long nanowires are thermodynamically unstable at high temperatures, since the nanowire length will almost always be much greater than its equilibrium solid radius. This has ramifications when considering device stability that utilise nanowires subjected to heating. We observed that for long, thin nanowires, the liquid-vapour interface can begin to destabilise even before the solid begins to neck. This implies ultra-long, thin nanowires will are particularly unstable at elevated temperatures and should be considered when constructing nanowire devices.

%% file: Stb_Conclusion.tex
\section{Conclusion}
We studied the stability of the solid in copper nanowires as they approach their melting temperature by perturbing a model describing interface kinetics and compared the results to MD simulations. The model found a stability criteria which dictates the preferred melting mode a nanowire will take. We found that longer nanowires are thermodynamically unstable, and will preferentially pinch-off and melt, indicating a melting mechanism driven by a PR type of instability. In shorter nanowires, the interface front moved towards the nanowire centre before the solid would breakup, indicating higher interface stability, with MD results in agreement with our model. Moreover, we proposed modes which destabilise the solid-liquid interface are proportional the nanowire length, which tells us that there are not preferred modes which destabilise the interface, in contrast to PR theory. Additionally, it was observed from the MD simulations that longer nanowires consistently have a melting temperature a few degrees below shorter nanowires, indicating the nanowire aspect ratio influences the preferred melting mode, and solid-liquid interfacial stability.

%% file: Acknowledgements.tex
\section{Acknowledgements}
The authors thank the MacDiarmid Institute for Advanced Materials and Nanotechnology for funding. The authors also wish to acknowledge the use of New Zealand eScience Infrastructure (NeSI) high performance computing facilities, consulting support and/or training services as part of this research. New Zealand's national facilities are provided by NeSI and funded jointly by NeSI's collaborator institutions and through the Ministry of Business, Innovation \& Employment's Research Infrastructure programme. URL https://www.nesi.org.nz.